

\documentclass[conference]{IEEEtran}
\IEEEoverridecommandlockouts

\usepackage{cite}
\usepackage{amsmath,amssymb,amsfonts}
\usepackage{algorithmic}
\usepackage{graphicx}
\usepackage{textcomp}
\usepackage{xcolor}
\usepackage{hyperref}

\def\BibTeX{{\rm B\kern-.05em{\sc i\kern-.025em b}\kern-.08em
    T\kern-.1667em\lower.7ex\hbox{E}\kern-.125emX}}

\begin{document}

\title{Spectral Bias Correction in PINNs for Myocardial Image Registration of Pathological Data}

\author{
    \IEEEauthorblockN{Bastien Carreon Baluyot}
    \IEEEauthorblockA{\textit{Electrical and Electronic Engineering} \\
    \textit{Imperial College London}\\
    London, UK \\
    bastien.baluyot20@alumni.imperial.ac.uk}
    
    \and
    \IEEEauthorblockN{Marta Varela\IEEEauthorrefmark{1}}
    \IEEEauthorblockA{\textit{National Heart and Lung Institute} \\
    \textit{Imperial College London}\\
    London, UK \\
    marta.varela@imperial.ac.uk}

    \and
    \IEEEauthorblockN{Chen Qin\IEEEauthorrefmark{1}}
    \IEEEauthorblockA{\textit{Electrical and Electronic Engineering} \\
    \textit{Imperial College London}\\
    London, UK \\
    c.qin15@imperial.ac.uk}

    \thanks{\IEEEauthorrefmark{1}These authors contributed equally to this work as senior authors.}
}

\maketitle

\begin{abstract}
Accurate myocardial image registration is essential for cardiac strain analysis and disease diagnosis. However, spectral bias in neural networks impedes modeling high-frequency deformations, producing inaccurate, biomechanically implausible results, particularly in pathological data. This paper addresses spectral bias in physics-informed neural networks (PINNs) by integrating Fourier Feature mappings and introducing modulation strategies into a PINN framework. Experiments on two distinct datasets demonstrate that the proposed methods enhance the PINN's ability to capture complex, high-frequency deformations in cardiomyopathies, achieving superior registration accuracy while maintaining biomechanical plausibility  ---  thus providing a foundation for scalable cardiac image registration and generalization across multiple patients and pathologies.
\end{abstract}


\section{Introduction}

Cardiovascular diseases (CVDs) are a major global health concern, responsible for millions of deaths each year \cite{WHO:21}. Accurate assessment of cardiac function is crucial for managing these conditions. Cardiac strain can be derived from cine-magnetic resonance imaging (cMRI), providing insights into the heart’s mechanical performance and aiding in diagnosing cardiomyopathies \cite{ScaBarBuc:17}. This analysis depends on precise image registration to align images throughout the cardiac cycle and calculate deformations.

However, predicting biomechanically realistic deformations is challenging due to the ill-posed nature of image registration and the absence of physically informed constraints in conventional and deep learning methods, often resulting in implausible deformations that compromise clinical applications.

Physics-informed neural networks (PINNs) address this challenge by incorporating physical laws into the learning process \cite{RaiPerKar:19}. Embedding myocardial incompressibility as a constraint \cite{Hum:02} improves clinical validity, making PINNs suitable for medical image registration. Recent work, such as the WarpPINN framework \cite{ArrMelUri:23}, leverages PINNs for cardiac image registration by introducing physics-informed biases using a neo-Hookean strain energy function as a regularizer on a discrete set of voxels.

However, a key challenge in using neural networks for image registration is spectral bias, which causes networks to favor low-frequency components over high-frequency ones \cite{Rah:19}. This bias can hinder the modeling of fine details and localized deformations, crucial for myocardial imaging, especially during rapid motion dynamics or pathological changes, leading to oversmoothing of deformation fields and potential loss of diagnostic information.

This work builds on existing techniques for implicit deformable image registration (IDIR) that address spectral bias through modulation and conditioning strategies to alleviate spectral bias \cite{SitMarBer:20, WolZwiBru:22, Zim:23, MehGhaBar:21}. Building upon these insights, this work evaluates the use of Fourier Feature mappings and proposes methods leveraging periodic activation functions and extended modulation techniques.

This paper aims to enhance myocardial image registration by addressing spectral bias in PINNs. The primary objectives and contributions are threefold:
\begin{enumerate}
    \item Demonstrating enhanced image registration capabilities of PINNs on pathological data by predicting biomechanically plausible deformations.
	\item Analyzing observations of spectral bias associated with cardiomyopathies.
    \item Proposing extended strategies to mitigate spectral bias in PINN-based image registration.
\end{enumerate}

\section{Background}

\subsection{Myocardial Image Registration}

Image registration aligns images by finding a transformation function $\boldsymbol{\varphi}$ that maps coordinates from a reference image $\boldsymbol{R}$ to a template image $\boldsymbol{T}$. The objective is to minimize the difference between the warped template image $\boldsymbol{T} \circ \boldsymbol{\varphi}$ and $\boldsymbol{R}$, as formulated in~\eqref{eq:Image_Reg_Formulation}.

\begin{equation}
    \min_{\boldsymbol{\varphi}} \; \mathcal{L}_{\text{similarity}}(\boldsymbol{R}, \boldsymbol{T} \circ \boldsymbol{\varphi}) + \mu \, \mathcal{L}_{\text{regularization}}(\boldsymbol{\varphi})
\label{eq:Image_Reg_Formulation}
\end{equation}

In the context of cardiac image registration, $\boldsymbol{R}$ and $\boldsymbol{T}$ are commonly taken to be the images at the end-diastolic (ED) and end-systolic (ES) phases of the cardiac cycle, respectively.

\subsection{Physics-Informed Neural Networks (PINNs)}

Physics-informed neural networks (PINNs) integrate physical laws into neural networks \cite{RaiPerKar:19}, constraining the solution space and enhancing interpretability. In myocardial image registration, PINNs incorporate cardiac tissue biomechanics. Enforcing near-incompressibility of the myocardium ensures predicted deformation fields are physically plausible \cite{ArrMelUri:23}, improving clinical validity over traditional data-driven methods lacking physical regularization.

\subsection{Spectral Bias in Neural Networks}

Spectral bias (or frequency bias) refers to the tendency of neural networks to learn low-frequency components faster than high-frequency ones \cite{Rah:19}, leading to oversmoothing.

In myocardial image registration, capturing intricate deformations  ---  especially in pathological cases  ---  requires learning high-frequency components corresponding to fine structural details and localized deformations. Spectral bias hinders this, causing an inaccurate representation of sharp transitions in the myocardium.

\section{Methods}

\subsection{WarpPINN}

This paper builds upon WarpPINN, a PINN-based framework for registering deformations of the left ventricle (LV) across the cardiac cycle \cite{ArrMelUri:23}. This PINN approach uses a multi-layer perceptron (MLP) to parameterize and learn a continuous deformation function.

\textbf{Architecture:} In this work, the WarpPINN model used employs an MLP with 5 fully-connected hidden layers (64 neurons each) using hyperbolic tangent ($\tanh$) activation functions. The input layer has 4 neurons for the 3D image $\boldsymbol{X}$ and time $t$, while the output layer has 3 neurons for the 3D non-rigid transformation $\boldsymbol{\varphi}$. Thus, the network represents a learned displacement field $\boldsymbol{u}(\boldsymbol{X}; t; \boldsymbol{\theta})$  ---  where $\boldsymbol{\theta}$ denotes the network parameters (weights and biases)  ---  such that $\boldsymbol{\varphi}(\boldsymbol{X}; \boldsymbol{\theta}) = \boldsymbol{X} + \boldsymbol{u}(\boldsymbol{X}; t; \boldsymbol{\theta})$.

\textbf{Physics-informed bias:} An observational bias is first introduced using a binary mask indicating whether a voxel is inside the LV. A learning bias is then incorporated into the loss function as in~\eqref{eq:WarpPINN_Loss_Eqn}.

\begin{equation}
    \mathcal{L}(\boldsymbol{\varphi}(\boldsymbol{X}; \boldsymbol{\theta}); \mu; \lambda) = |\boldsymbol{R} - \boldsymbol{T} \circ \boldsymbol{\varphi}|_{1} + \mu \mathcal{R}(\boldsymbol{\varphi}; \lambda)
\label{eq:WarpPINN_Loss_Eqn}
\end{equation}

The first term is an image similarity metric, and the second term is a regularization term for the predicted deformation. $\mathcal{R}(\boldsymbol{\varphi}; \lambda)$ is chosen to be the neo-Hookean hyperelastic strain energy function: $W(\boldsymbol{\varphi}; \lambda) = \text{Tr}(\boldsymbol{C}) - 3 - 2 \log(\boldsymbol{J}) + \lambda (\boldsymbol{J} - 1)^2$, where \(\boldsymbol{F} = \frac{\partial \boldsymbol{\varphi}}{\partial \boldsymbol{X}}\) is the deformation gradient, \(\boldsymbol{C} = \boldsymbol{F}^T \boldsymbol{F}\) is the right Cauchy-Green tensor, and \(\boldsymbol{J} = \det(\boldsymbol{F})\) is the determinant of the deformation gradient. 

The $\lambda(\boldsymbol{J} - 1)^{2}$ term penalizes changes in volume, being zero when incompressibility is satisfied. Thus, myocardial (quasi)incompressibility is enforced by applying $\mathcal{R}(\boldsymbol{\varphi}; \lambda)$ only to voxels inside the LV, as indicated by the binary mask.

\subsection{Fourier Feature Mapping (FFM)}

Fourier Feature (FF) encoding addresses spectral bias by mapping input coordinates to a higher dimensional, higher frequency domain \cite{Tan:20}. This can be formulated with the original input image $\boldsymbol{X}$. Defining the matrix \(\boldsymbol{B} \in \mathbb{R}^{m \times d}\), where \(m\) is a positive integer, \(d\) is the input dimension, and each entry \(\boldsymbol{B}_{i,j}\) is sampled independently from a normal distribution \(\mathcal{N}(0, \sigma^2)\), the FF mapping $\boldsymbol{\gamma}(\boldsymbol{X})$ is thus defined as: $\boldsymbol{\gamma}(\boldsymbol{X}) = \begin{bmatrix} \cos(\boldsymbol{B}\boldsymbol{X}) & \sin(\boldsymbol{B}\boldsymbol{X}) \end{bmatrix}^T$.

\subsection{Sinusoidal Representation Networks (SIRENs)}

Sinusoidal representation networks (SIRENs) use a sine activation function $\psi_{SIREN}(x) = \sin(\omega_0 \: x)$, where $\omega_{0}$ controls the periodicity. Periodic activation functions alleviate spectral bias by preventing exploding and vanishing gradients \cite{MehGhaBar:21}, enabling training on a broader range of image pairs. In this work, WarpPINN is adapted as a SIREN, using sinusoidal activation functions and weight initializations as in \cite{SitMarBer:20}.

\begin{figure*}[ht]
\centering
\includegraphics[width=1\textwidth]{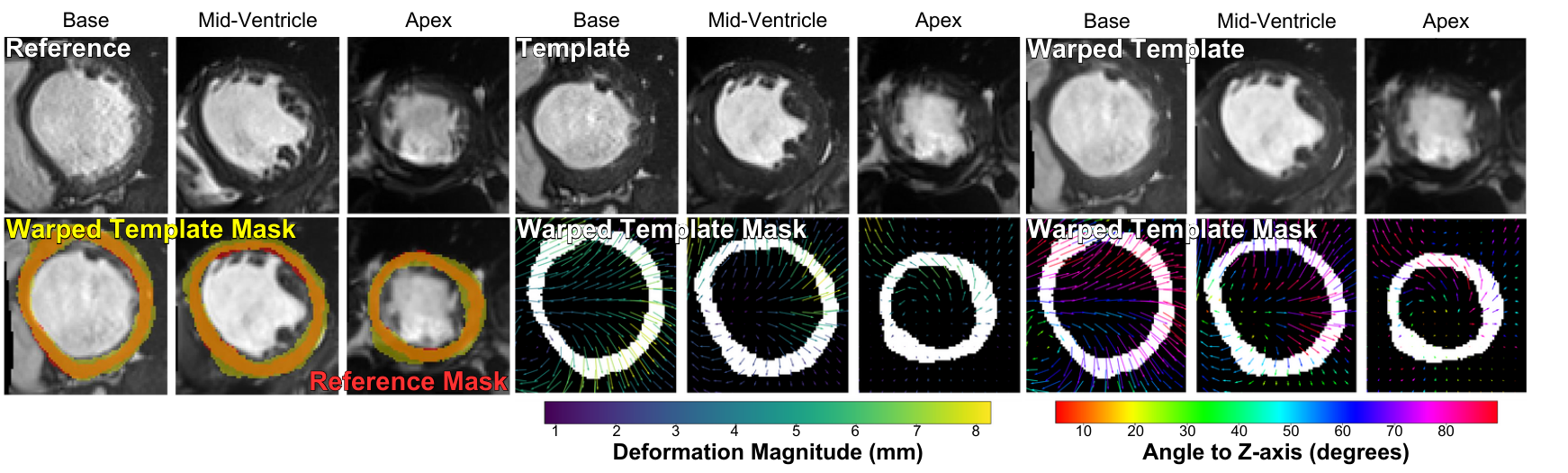}
\caption{Visualization of Reference ($\boldsymbol{R}$), Template ($\boldsymbol{T}$) and Warped Template ($\boldsymbol{T} \circ \boldsymbol{\varphi}$) images (and segmentation masks) for Patient 50 (MINF) from the ACDC dataset. Visualizations are taken at the three representative slices (basal, mid-ventricular, and apical). Glyphs are shown to visualize the magnitude and direction of the learned deformation field ($\boldsymbol{\varphi}$).}
\label{fig:ACDC_Patient50_Registration_Visualization}
\end{figure*}

Building upon the SIREN approach, this work introduces modulation strategies inspired by prior work on generalizing IDIR \cite{Zim:23}. These strategies are extended to include Phase Shift Keying (PSK) and Quadrature Phase Shift Keying (QPSK). Instead of using latent space vectors for modulation as in previous works \cite{AmiLudLi:22}, this paper's approach leverages the FF representations of the input data as modulation inputs to effectively encode high-frequency image details, enhancing registration accuracy and generalizability across different image pairs. The following modulation strategies are investigated:

\begin{enumerate}
    \item \textbf{Fourier Feature SIREN} (FF-S): The Fourier Feature representation of the image $\boldsymbol{\gamma}(\boldsymbol{X})$ is fed directly as an input to the SIREN.
    
    \item \textbf{Amplitude Modulation} (AM): The activation function of the SIREN is multiplied by a scalar value $\alpha$.
    \begin{equation}
        \psi_{AM}(x) = \alpha \sin(x)
    \end{equation}

    \item \textbf{Phase Shift Keying} (PSK): The activation function of the SIREN is phase-shifted by an angle $\phi$.
    \begin{equation}
        \psi_{PSK}(x) = \sin(x + \phi)
    \end{equation}

    \item \textbf{Quadrature Phase Shift Keying} (QPSK): The activation function of the SIREN is a combination of sine and cosine functions, each phase-shifted by an angle $\phi$.
    \begin{equation}
        \psi_{QPSK}(x) = \sin(x + \phi) + \cos(x + \phi)
    \end{equation}
\end{enumerate}

The proposed techniques enable the SIREN to have a modulated response to varying image pairs (across the cardiac cycle) by dynamically adjusting to the input data, parameterized as FF representations. This is further enhanced through an adaptive frequency response approach to computing the key modulation parameters, $\alpha$ and $\phi$, as follows below.

\begin{quote}
    Let $\mathcal{E}$ be the total energy of the input signal.    
\end{quote}

\begin{equation}
    \mathcal{E} = \sum_{i=1}^{m} \left( \cos(\boldsymbol{BX})_{i} + \sin(\boldsymbol{BX})_{i} \right)^2
\end{equation}

\begin{equation}
    \alpha = \frac{1}{1 + \exp(- \mathcal{E})} = \text{sigmoid}(\mathcal{E})
\end{equation}

\begin{equation}
    \phi = \text{atan2} \left( \sum_{i=1}^{m} \sin \left( \boldsymbol{B} \boldsymbol{X} \right)_{i},  \sum_{i=1}^{m} \cos \left( \boldsymbol{B} \boldsymbol{X} \right)_{i} \right)
\end{equation}

\begin{quote}
    where $m$ is the number of features (half the dimensionality of $\boldsymbol{\gamma}(\boldsymbol{X})$) and $\text{atan2}(y, x)$ is the two-argument arctangent function.
\end{quote}

\subsection{Datasets}

The experiments outlined are carried out using two publicly available cardiac cine-MRI datasets:

\begin{enumerate}

    \item \textbf{CMAC:} The Cardiac Motion Analysis Challenge dataset includes cMRI scans of 15 healthy volunteers with ground truth coordinates of 12 anatomical landmarks and a segmentation mask of the LV at the ED phase \cite{Tob:13}.

    \item \textbf{ACDC:} The Automated Cardiac Diagnosis Challenge dataset comprises cMRI sequences for 150 patients from ED to ES. Ground truth myocardium segmentation masks at ED and ES are provided; patients are divided equally into 5 subgroups\footnote{Considered disease groups in this study are: healthy (\textbf{NOR}); dilated cardiomyopathy (\textbf{DCM}); hypertrophic cardiomyopathy (\textbf{HCM}); and patients with previous myocardial infarction (\textbf{MINF})} (4 pathological, 1 healthy) \cite{Ber:18}.

\end{enumerate}

\section{Experiments and Results}

The code used will be made publicly available here\footnote{https://github.com/BastienB04/Myo-PINN-SIREN}.

\subsection{PINN-based Registration on ACDC Dataset}

\textbf{Aim:} To evaluate the image registration performance of WarpPINN on pathological data.

\textbf{Training Regime:} For each patient in the ACDC dataset, a vanilla WarpPINN model (without FFM) is trained using the ED image as $\boldsymbol{R}$ and a randomly chosen frame as $\boldsymbol{T}$ in each iteration. Training is done with an ADAM optimizer and a learning rate of $10^{-3}$ for $3\times10^{5}$ iterations. The weights are initialized with Xavier initialization and the biases as zero. The $\lambda$ and $\mu$ hyperparameters are tuned with a Bayesian optimization approach to maximize image similarity whilst adhering to the volume preservation constraint.

\textbf{Evaluation Measure:} Image registration performance is assessed using the Dice Similarity Coefficient (DSC) and Mean Contour Distance (MCD), measuring the overlap between warped template and reference segmentation masks. Evaluations are performed at three representative short-axis slices (basal, mid-ventricular, and apical), defined at 25\%, 50\%, and 75\% of the LV length, respectively (visualized in Fig.~\ref{fig:ACDC_Patient50_Registration_Visualization}). Volume preservation is measured by the determinant of the Jacobian $\det(\boldsymbol{J})$ across the myocardium, being 1 when volume is perfectly conserved (incompressible behavior).

\textbf{Benchmark Methods:} Registration performance is compared to five other methodologies. Classical methods include Free-Form Deformation with Volumetric Preservation (FFD-VP) \cite{RohMau:01} and Diffeomorphic Demons (dDemons) \cite{VerPenPer:07}. DL-based methods include Motion-Net (a CNN-based approach) \cite{Qin:18}, a Biomechanics-Informed Neural Network (BINN) \cite{QinWanChe:20} approach, and a Biomechanics-Informed Generative Model (BIGM-VAE) \cite{QinWanChe:23} that utilizes a pre-trained temporal VAE. 

Disease-aggregated results for the comparison of WarpPINN and BIGM-VAE are presented in Table~\ref{tab:WarpPINN_vs_BIGMVAE} and comparative results across the entire ACDC cohort are presented in Table~\ref{tab:ACDC_Performance} as mean(standard deviation), where \textbf{bold} and \underline{underlined} results indicate the best and second best performance per metric, respectively. Registration performance data for the benchmark methodologies are provided in \cite{QinWanChe:23}.

\begin{table}[htbp]
\centering
\caption{Comparison of image registration performance of WarpPINN and BIGM-VAE methods on ACDC dataset by disease group}
\label{tab:WarpPINN_vs_BIGMVAE}
\begin{tabular}{lllll}
\hline \hline
Disease & Method   & MCD                   & DSC                   & $||\boldsymbol{J}|-1|$ \\ \hline
NOR     & WarpPINN & \textbf{1.256(0.076)} & \textbf{0.727(0.017)} & 0.164(0.015)           \\
        & BIGM-VAE & 3.123(0.939)          & 0.704(0.076)          & \textbf{0.140(0.037)}  \\ \hline
DCM     & WarpPINN & \textbf{0.912(0.145)} & 0.793(0.018)          & 0.085(0.012)           \\
        & BIGM-VAE & 1.260(0.221)          & \textbf{0.832(0.021)} & \textbf{0.084(0.021)}  \\ \hline
HCM     & WarpPINN & \textbf{1.173(0.259)} & \textbf{0.783(0.045)} & \textbf{0.113(0.015)}  \\
        & BIGM-VAE & 3.398(0.934)          & 0.764(0.047)          & 0.149(0.025)           \\ \hline
MINF    & WarpPINN & 1.655(0.329)          & 0.725(0.029)          & 0.116(0.024)           \\
        & BIGM-VAE & \textbf{1.549(0.196)} & \textbf{0.801(0.038)} & \textbf{0.116(0.017)}  \\ \hline \hline
\end{tabular}
\end{table}

\begin{table*}[ht]
\centering
\setlength{\tabcolsep}{4pt}
\caption{Comparison of image registration performance of all considered methodologies on the ACDC dataset}
\label{tab:ACDC_Performance}
\begin{tabular}{llllllllll}
\hline\hline
\textbf{Method} & \multicolumn{3}{c}{\textbf{Base}} 
                & \multicolumn{3}{c}{\textbf{Mid-Ventricle}} 
                & \multicolumn{3}{c}{\textbf{Apex}} \\
\cline{2-4}\cline{5-7}\cline{8-10}
                & MCD               & DSC               & $||\boldsymbol{J}|-1|$
                & MCD               & DSC               & $||\boldsymbol{J}|-1|$
                & MCD               & DSC               & $||\boldsymbol{J}|-1|$ \\
\hline
FFD-VP          & 2.925 (1.328)     & 0.731 (0.120)     & 0.135
                & 2.442 (1.352)     & 0.756 (0.088)     & 0.147
                & 2.784 (1.648)     & 0.684 (0.143)     & 0.148 \\
                &                   &                   & (0.052)
                &                   &                   & (0.072)
                &                   &                   & (0.075) \\
\hline
dDemons         & 2.295 (1.174)     & 0.778 (0.102)     & \underline{0.132}
                & 1.993 (1.112)     & \underline{0.788 (0.075)}  & 0.141
                & 2.437 (1.453)     & 0.707 (0.128)     & \underline{0.132} \\
                &                   &                   & \underline{(0.031)}
                &                   &                   & (0.048)
                &                   &                   & \underline{(0.043)} \\
\hline
Motion-Net      & 2.814 (1.236)     & 0.751 (0.123)     & 0.167
                & 2.799 (1.007)     & 0.745 (0.105)     & 0.167
                & 2.853 (1.301)     & 0.656 (0.142)     & 0.171 \\
                &                   &                   & (0.049)
                &                   &                   & (0.056)
                &                   &                   & (0.060) \\
\hline
BINN            & 2.229 (0.860)     & \underline{0.789 (0.091)}   & 0.161
                & 2.210 (0.918)     & 0.783 (0.097)     & 0.148
                & 2.450 (1.253)     & 0.707 (0.147)     & 0.158 \\
                &                   &                                 & (0.063)
                &                   &                                 & (0.056)
                &                   &                                 & (0.081) \\
\hline
BIGM-VAE        & \underline{1.660 (0.671)} & \textbf{0.829 (0.067)} & 0.138
                & \underline{1.646 (0.971)} & \textbf{0.818 (0.055)} & \underline{0.134}
                & \underline{2.101 (1.312)} & \underline{0.731 (0.137)} & 0.134 \\
                &                           &                         & (0.060)
                &                           &                         & \underline{(0.067)}
                &                           &                         & (0.091) \\
\hline
WarpPINN        & \textbf{1.106 (1.072)} & 0.781 (0.095)       & \textbf{0.144}
                & \textbf{1.236 (0.932)} & 0.759 (0.082)       & \textbf{0.126}
                & \textbf{1.374 (1.439)} & \textbf{0.733 (0.085)} & \textbf{0.089} \\
                &                         &                     & \textbf{(0.062)}
                &                         &                     & \textbf{(0.050)}
                &                         &                     & \textbf{(0.042)} \\
\hline\hline
\end{tabular}
\end{table*}

\begin{figure*}[ht]
\centering
\includegraphics[width=0.82\textwidth]{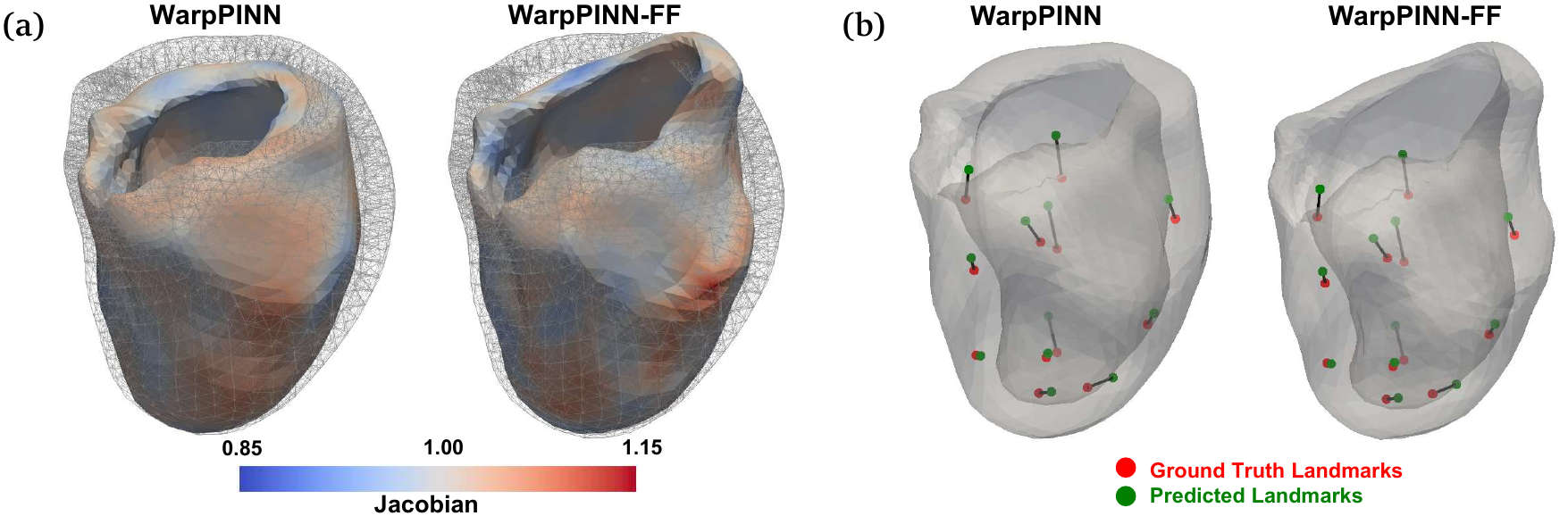}
\caption{3D visualization of the predicted deformation of the LV at ES phase for one healthy volunteer from the CMAC dataset for the WarpPINN and WarpPINN-FF models with $\lambda = 10^{5}$, $\mu = 5 \times 10^{-6}$, and $\sigma = 1$ (for WarpPINN-FF). (a) Surface colormap of the Jacobian with the wireframe showing the ground truth surface mesh at ED. (b) Predicted (green) vs. ground truth (red) landmark coordinates at ES phase, with black lines indicating tracking error.}
\label{fig:CMAC_Volunteer1_Registration_Visualization}
\end{figure*}

\subsection{Spectral Bias Correction on CMAC Dataset}

The CMAC dataset, featuring a 3D ground truth surface mesh, is used to compare WarpPINN and WarpPINN-FF (using FFM), allowing visualization of localized deformations.

\textbf{Training Regime:} The WarpPINN and WarpPINN-FF models are trained with an identical training regime as that in the ACDC experiment. However, the $\lambda$ and $\mu$ parameters are fixed to $10^{5}$ and $5 \times 10^{-6}$, respectively, as in \cite{ArrMelUri:23} (optimal hyperparameters for CMAC data). For experiments with WarpPINN-FF, the parameters $m=8$ and $\sigma=1$ are chosen.

\textbf{Evaluation Measure:} After training, the 12 ground truth landmarks at ED are input to the trained network to predict their locations at ES by warping with the learned deformation $\boldsymbol{\varphi}$. Registration performance is evaluated by measuring the Euclidean distance between predicted and ground truth landmark coordinates (Landmark Tracking Error). This is shown in Fig.~\ref{fig:CMAC_Volunteer1_Registration_Visualization}, alongside visualizations of the predicted deformations and the Jacobian measure across the myocardium surface.

\begin{figure}[h]
\centering
\includegraphics[width=1\columnwidth]{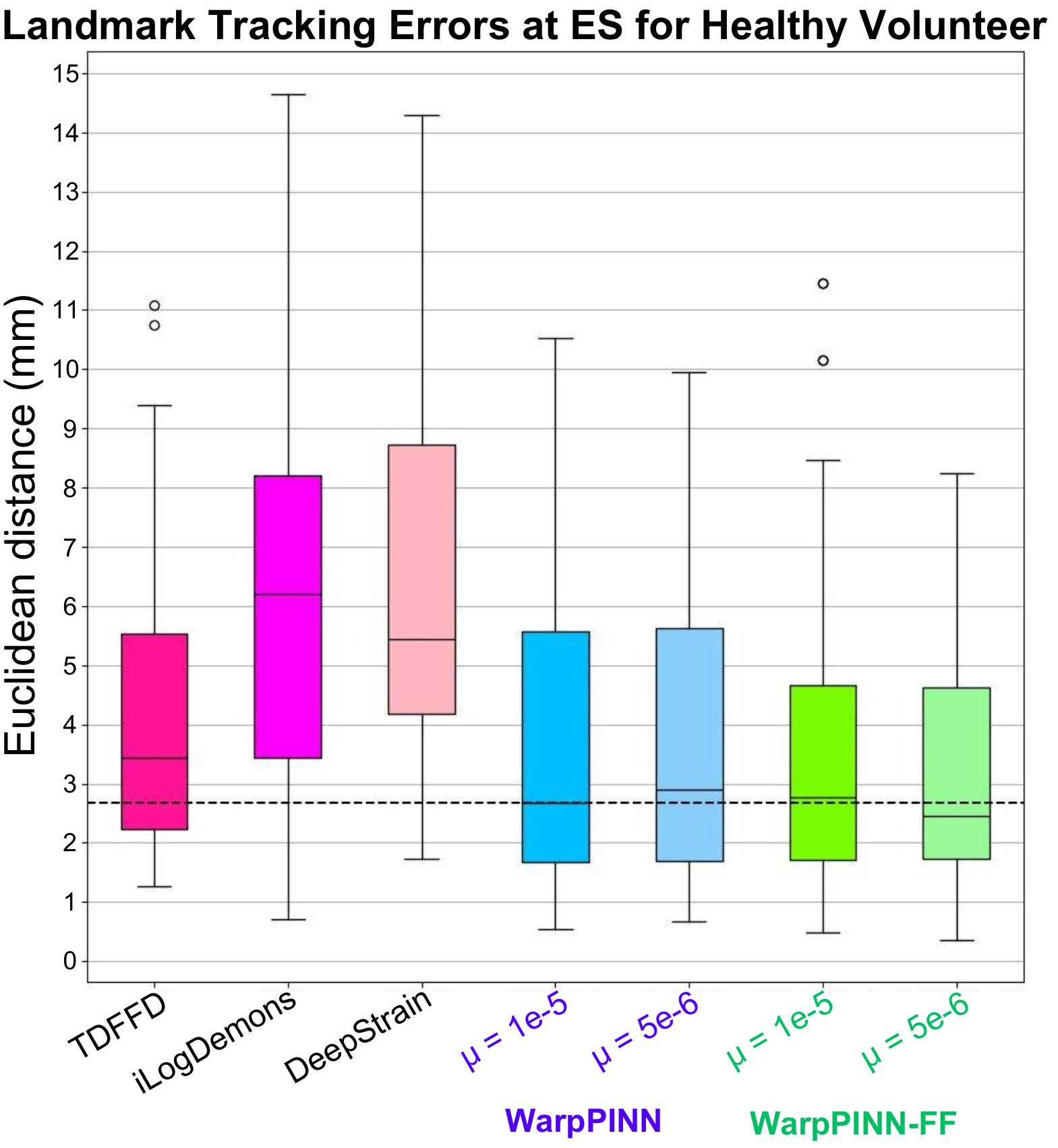}
\caption{Box plot of landmark tracking errors of various image registration methodologies for registration on one healthy volunteer from the CMAC dataset. WarpPINN-FF is evaluated with hyperparameter $\sigma = 1$. Registration performance data for the benchmark methodologies are provided in \cite{ArrMelUri:23}.}
\label{fig:CMAC_Landmark_Tracking_Error_Box_Plot}
\end{figure}

\textbf{Benchmark Methods:} The performance of three different benchmark image registration methods is  used for comparison. These include classical image registration algorithms, namely Temporal Diffeomorphic Free Form Deformation (TDFFD) \cite{DeC:12} and iLogDemons \cite{ManPenSer:11}, as well as a CNN-based approach, DeepStrain \cite{Mor:21}. Comparative results are presented in Fig.~\ref{fig:CMAC_Landmark_Tracking_Error_Box_Plot}.

\subsection{Evaluation of SIRENs}

To evaluate the performance of the proposed SIREN models on pathological data, each of the proposed models is trained on a subset of the ACDC dataset, comprising 10 patients from each disease group, with the same training regime as the previous ACDC experiment and the $\lambda$ and $\mu$ hyperparameters being optimized accordingly. A baseline vanilla WarpPINN model with the same optimal hyperparameters is used for comparison. Results are presented in Table~\ref{tab:SIREN_ACDC}.

\section{Discussions of Results}

\subsection{PINNs on Pathological Data}

WarpPINN demonstrates strong performance on the ACDC dataset across various cardiac pathologies. As shown in Fig.~\ref{fig:ACDC_Patient50_Registration_Visualization}, the ED and ES masks are effectively aligned while satisfying myocardial incompressibility.

Across the ACDC dataset, WarpPINN outperforms all benchmarks in MCD and achieves comparable DSC scores (Table~\ref{tab:ACDC_Performance}). Lower MCD indicates more precise myocardial contour alignment, likely due to continuous signal parameterization that enables finer deformation modeling. Notably, WarpPINN also exhibits the lowest $||\boldsymbol{J}|-1|$ across all slice levels, indicating superior volume preservation, suggesting effective learning of accurate deformation fields while maintaining myocardial incompressibility.

\begin{table}[h]
\centering
\caption{Registration performance of proposed SIRENs on ACDC data}
\label{tab:SIREN_ACDC}
\begin{tabular}{lllll}
\hline\hline
\textbf{Method} & \multicolumn{2}{c}{\textbf{NOR}} 
                & \multicolumn{2}{c}{\textbf{DCM}} \\
\cline{2-5}
                & DSC                 & $||\boldsymbol{J}|-1|$ 
                & DSC                 & $||\boldsymbol{J}|-1|$ \\
\hline
FF-S            & \textbf{0.783(0.114)}    & 0.209(0.046)
                & \underline{0.785(0.090)}  & \textbf{0.048(0.017)} \\
AM              & 0.767(0.115)             & \underline{0.198(0.039)}
                & 0.782(0.150)             & 0.110(0.014)          \\
PSK             & 0.733(0.116)             & 0.210(0.033)
                & 0.766(0.112)             & 0.071(0.026)          \\
QPSK            & \underline{0.772(0.115)} & 0.203(0.032)
                & 0.758(0.142)             & \underline{0.069(0.030)} \\
Vanilla         & 0.726(0.092)             & \textbf{0.165(0.029)}
                & \textbf{0.796(0.059)}    & 0.116(0.077)          \\
\hline
\textbf{Method} & \multicolumn{2}{c}{\textbf{HCM}} 
                & \multicolumn{2}{c}{\textbf{MINF}} \\
\cline{2-5}
                & DSC                 & $||\boldsymbol{J}|-1|$ 
                & DSC                 & $||\boldsymbol{J}|-1|$ \\
\hline
FF-S            & \underline{0.907(0.073)} & 0.153(0.014)
                & \textbf{0.850(0.083)}    & 0.182(0.024)          \\
AM              & 0.893(0.041)             & \underline{0.132(0.006)}
                & \underline{0.833(0.052)} & 0.159(0.030)          \\
PSK             & 0.895(0.074)             & 0.143(0.024)
                & 0.815(0.047)             & \underline{0.155(0.013)} \\
QPSK            & \textbf{0.913(0.068)}    & 0.169(0.012)
                & 0.817(0.090)             & 0.212(0.008)          \\
Vanilla         & 0.788(0.073)             & \textbf{0.113(0.002)}
                & 0.797(0.028)             & \textbf{0.139(0.046)} \\
\hline\hline
\end{tabular}
\end{table}
 
Across different cardiac conditions (Table~\ref{tab:WarpPINN_vs_BIGMVAE}), WarpPINN tends to perform better in patients with hypertrophic cardiomyopathy (HCM) and normal (NOR) hearts, in terms of image similarity. HCM features thickened myocardium and potentially more localized, complex deformation patterns due to asymmetric hypertrophy, while NOR represents healthy hearts with intricate motion dynamics \cite{Ell:08}. WarpPINN's higher performance in these particular groups could suggest its effectiveness in capturing high-frequency deformations.
 
Conversely, BIGM-VAE performs better in dilated cardiomyopathy (DCM) and myocardial infarction (MINF) patients. DCM involves global ventricular dilation with reduced motion, leading to smoother, low-frequency deformations. MINF is associated with scar tissue causing akinesia (no movement) or dyskinesia (abnormal movement), and thus creating spatial heterogeneity in deformations \cite{Ell:08}. The superior performance of BIGM-VAE in these conditions indicates that its architecture may be adept at capturing the specific deformation patterns associated with these pathologies.

These findings illustrate the complex relationship between cardiac pathology, deformation patterns, and registration performance. WarpPINN’s physics-informed approach is beneficial in HCM and NOR data, while other architectures may be more suitable in DCM and MINF data. This highlights the need to tailor registration methods to the specific characteristics of each pathology. To enhance registration performance across diverse cardiac conditions, integrating spectral bias mitigation techniques with pathology-specific deformation considerations is crucial. This motivates further exploration of advanced methods in the subsequent section.

\subsection{Alleviating Spectral Bias}

Experiments on the CMAC dataset reveal that integrating FFM into WarpPINN improves performance. As shown in Fig.~\ref{fig:CMAC_Volunteer1_Registration_Visualization}, WarpPINN-FF predicts sharper, more localized deformations than the vanilla WarpPINN.

Figure~\ref{fig:CMAC_Landmark_Tracking_Error_Box_Plot} indicates that WarpPINN models achieve lower landmark tracking errors than other methods on CMAC data, demonstrating superior registration performance. Although both models display similar accuracy, WarpPINN-FF achieves more consistent results (lower variance) than the vanilla model.

Addressing spectral bias further, results in Table~\ref{tab:SIREN_ACDC} indicate that the proposed SIREN methods outperform vanilla WarpPINN in DSC for most disease groups while maintaining comparable volume preservation, particularly in the HCM and MINF groups, where high-frequency deformations are common. However, in the DCM group, SIREN methods slightly underperform, suggesting that the vanilla model’s low-frequency bias is beneficial for smoother deformations.

These findings demonstrate that incorporating FFM and the proposed periodic activation functions into WarpPINN effectively addresses spectral bias, improving the network’s capacity to capture complex, high-frequency deformations essential for accurate myocardial image registration in pathological conditions.

\subsection{Limitations and Future Work}

Despite promising results, several limitations exist. The biomechanical model employed in the WarpPINN studies is relatively simplistic, prompting further investigation on various model formulations for pathological datasets presenting abnormal mechanics.

Additionally, WarpPINN-FF is evaluated only on the CMAC dataset, which has ground truth surface meshes for deformation visualization. The absence of these meshes in the ACDC dataset limits evaluations on pathological data. Future efforts could involve ACDC mesh generation as in \cite{JoyBuoSto:22}, enabling cross-dataset validation and better assessment of FFM.

The proposed SIREN approaches are also tested on a limited subset of the ACDC dataset, providing only preliminary insights into their ability to alleviate spectral bias. Expanding experiments to include larger, more diverse patient cohorts and other datasets  ---  like M\&Ms \cite{Cam:21} and UKBB \cite{Pet:17}  ---  would enhance the validity and generalizability of the findings. Addressing computational constraints could involve investigating more efficient training strategies or developing generalized models beyond pairwise optimization. 

Moreover, while the SIREN methods show promise, they depend on the proposed adaptive frequency response approach for computing modulation parameters. Exploring alternatives like learned embeddings \cite{ParFloStr:19} or attention mechanisms \cite{Vas:17} could enhance modulation further. Overall, these SIREN-based techniques may help generalize PINNs across multiple patients without relying on pairwise optimization, such as through the use of an encoder to condition modulation based on latent space vectors, as done in \cite{Zim:23}. This would effectively address another key limitation of current registration methods.

\section{Conclusion}

This study demonstrates that PINNs enhance myocardial image registration by predicting biomechanically plausible transformations, outperforming conventional and deep learning methods on the CMAC and ACDC datasets. Trends related to spectral bias in registering pathological cMRI data highlight the need to address this challenge. Utilizing FFM and the proposed SIREN approaches enhances PINNs' ability to register high-frequency deformations in cardiomyopathies. 

The success of the proposed SIREN methods provides a foundation for future research on generalizing PINN-based registration models across various patients and pathologies without pairwise optimization, enabling efficient, scalable myocardial image registration, potentially improving cardiac motion analysis and patient care.

\end{document}